\documentclass[sn-mathphys,Numbered]{sn-jnl}

\usepackage{graphicx}%
\usepackage{multirow}%
\usepackage{amsmath,amssymb,amsfonts}%
\usepackage{amsthm}%
\usepackage{mathrsfs}%
\usepackage[title]{appendix}%
\usepackage{xcolor}%
\usepackage{textcomp}%
\usepackage{manyfoot}%
\usepackage{booktabs}%
\usepackage{algorithm}%
\usepackage{algorithmicx}%
\usepackage{algpseudocode}%
\usepackage{listings}%
\usepackage{setspace}

\theoremstyle{thmstyleone}%
\theoremstyle{thmstyletwo}%
\theoremstyle{thmstylethree}%
\begin{document}

\title{The intricate path of energy conservation in Quantum Mechanics: exploring Coherent Population Return and laser-matter interaction}

\author{\fnm{Álvaro} \sur{Peralta Conde}}\email{alvaro.peralta@unir.net}

\affil{Universidad Internacional de la Rioja. UNIR. www.unir.net. Spain}

\abstract{This manuscript explores how a seemingly straightforward inquiry, emerging from the widely accepted semiclassical description of laser-matter interaction, concretely from a well-knows adiabatic technique as Coherent Population Return (CPR), can pose a challenge to our comprehension of a fundamental principle in Physics like energy conservation. Throughout our investigation to resolve this apparent paradox, we have delved into fundamental concepts, thereby deepening our understanding of the aspects inherent to the formalism of Quantum Mechanics. We emphasize that the significance of attaining a conclusive answer extends beyond the answer itself, encompassing the illuminating journey undertaken to reach it. Consequently, our work holds educational value as it aims to foster a deeper understanding of the phenomenon by elucidating the process employed. This approach not only aids students in grasping the subject matter but also enhances our own understanding of laser-matter interaction and the counterintuitive phenomena that continuously defied our understanding of Quantum Mechanics.}

\maketitle

\section{Introduction}\label{Introduction}
When it comes to teaching Science, we often make the mistake of presenting facts and ideas in isolation, without adequately considering the historical and contextual development behind them. Additionally, we fail to emphasize the interconnectedness of scientific concepts and their implications across different fields. As a result, Science is often taught as a collection of disjointed pieces, neglecting the inherent relationships and connections between scientific disciplines. One prime example that exemplifies this issue is the gap between mathematical concepts and their applications, such as the Algebra of matrices and Quantum Mechanics, which we will delve into further.

Regrettably, this approach to teaching is pervasive in educational institutions, despite its significant flaws. While it may work for students who already possess a strong motivation for Science, it tends to discourage others and dampen their interest in the subject. Moreover, recent advancements in educational theory have shown that the context in which concepts are presented plays a critical role in understanding them. Furthermore, if educators are able to actively engage students in the learning process, allowing them to take on a protagonistic role rather than a passive one, the chances of fostering significant and enduring learning increase substantially.

As an illustrative example, let us briefly delve into Constructivism, one of the most influential modern educational theories which is widely accepted across various educational systems, especially in primary and secondary schools (see for example \cite{Bruner96, Dewey97, Ausubel00} and references therein). Constructivism posits that passive acquisition of knowledge, through direct transmission, is significantly less effective from an educational standpoint compared to situations where students actively participate in the process, taking ownership of constructing their own knowledge. When students learn a concept, they construct a new understanding and knowledge based on their experiences and social context. Another key tenet of this theory is the recognition that newly acquired knowledge is not randomly integrated into a student’s cognitive structure but rather connected to prior knowledge and existing information. Hence, the generation of knowledge must be understood as an inherently active process. This stands in stark contrast to prevalent teaching methodologies in most universities, where knowledge is typically transmitted directly from teachers to learners, with the latter often seen as passive vessels to be filled with information.

Another aspect related to the aforementioned issues is the tendency to present scientific concepts as closed and monolithic constructions, avoiding any discussion about the various paths taken before arriving at the right one. This approach portrays Science as a linear discipline, disregarding the iterative and constructive nature of its development. For instance, it is not uncommon to hear phrases like "Newton invented differential calculus" or "Einstein invented general relativity," which not only overlook the contextual nuances surrounding these breakthroughs, but also undermine the significant contributions of other scientists to these advancements. It is well-known that Einstein’s groundbreaking development of General Relativity would not have been possible without the preceding studies of non-Euclidean geometry by Gauss, Riemann, and Lobachevsky among others.

Furthermore, aligning with Constructivism, when learners produce new knowledge, they are essentially exploring uncharted territories, which inherently opens the door to making mistakes and being wrong. In other words, errors are an integral part of the knowledge generation process. The development of any scientific theory serves as a clear example of this (see for example the foundational history of Quantum Mechanics \cite{ron01}). However, this clashes with the prevailing educational environment where mistakes are typically penalized, disregarding their immense educational potential. This issue extends beyond education and permeates researchers as well. Our current view of Science often considers negative results as inconclusive and, therefore, unpublishable. Thankfully, in recent years, efforts have been made to challenge this inertia. For instance, some journals now encourage authors to publish negative data alongside positive findings, recognizing the value of the former as complementary information.

With these ideas in mind, we present a story of a discussion filled with mistakes until the right answer was eventually found. It all began with a seemingly simple question that shook our understanding of the underlying Science, necessitating a revisiting of the most fundamental concepts. This process took considerable time and involved countless discussions where numerous mistakes, incorrect interpretations, and even nonsensical theories emerged. Eventually, one day, the breakthrough idea emerged, shedding light on the problem at hand. In the following sections, we will recount the events exactly as they unfolded. We believe that uncovering the scientific process in this manner can be immensely helpful for future students and possess undeniable educational value. We failed numerous times throughout this journey, but isn’t this precisely how Science progresses?

\section{The problem}\label{The problem}

The ability to control the inherent coherent nature of laser-matter interaction has proven to be a very valuable tool to steer population distributions, generate coherent superpositions of quantum states, support spectroscopical investigations, manipulate linear or nonlinear optical properties, and control photofragmentation channels. Techniques like Stimulated Raman Adiabatic Passage (STIRAP) \cite{Bergmann98}, Electromagnetic Induced Transparency (EIT) \cite{Harris97, Marangos98}, Retroreflection Induced Adiabatic Passage (RIBAP) \cite{Yatsenko03, Peralta05}, or Coherent Population Return (CPR) \cite{Vitanov01, Peralta06} have been successfully applied presenting a great robustness with respect to the experimental parameters. Let us focus on CPR. This coherent technique allows a non-resonant laser pulse to produce a non-permanent transfer of population from a ground state to a target excited state. The transfer of population is non-permanent as the population completely returns to the ground state once the excitation pulse ceases. This transient nature of the population transfer distinguishes CPR from other processes where the population may remain in the excited state even after the excitation pulse is finished. Also, it is important to note that this phenomenon has no classical analogy; it is a pure quantum mechanical effect. Additionally, a notable characteristic of CPR is the ability to employ a significantly large laser detuning. In fact, for the dynamics of the process to be adiabatic, the detuning must be larger than the laser bandwidth. This will be discussed in detail in the following sections.

In this scenario, let’s consider the introduction of a second laser to probe the transiently excited population, for example, through photoionization. This leads to several intriguing questions. How is the transfer of population achieved when the excitation laser is significantly detuned from resonance? How is ionization produced when the absorbed photon energy can be smaller than the system’s ionization potential? And how is energy conserved in the process? This situation somewhat resembles the quantum tunneling effect, where the wavefunction partially resides beneath a potential energy barrier while crossing it. Interrogating the wavefunction’s energy in such circumstances is not a trivial exercise. These thought-provoking questions have been the primary motivation for this research, and we will endeavor to address them in the subsequent investigation.

We believe that this work can be implemented within the framework of Problem-Based Learning or Project-Based Learning, which are student-centered educational approaches that emphasize active engagement in solving real-world problems or participating in real-world projects to develop knowledge and skills. The concepts and mathematics involved in this study are accessible to university-level students. These methodologies promote critical thinking, problem-solving skills, and the application of knowledge in practical contexts. Relevant literature on these educational approaches includes references such as \cite{Savery95, Hmelo04, Dolmans16}, for Problem-Based Learning, and references such as the \cite{Buck13, Krajcik06, Larmer15}, for Project-Based Learning.

\section{A two-state system interacting with a laser radiation}
For coherent interactions, the description of the population dynamics is provided by the time dependent Schr\"{o}dinger equation 
\begin{equation}
\label{timedepsch} 
i\hbar\frac{\partial\Psi(t)}{\partial t}=H(t)\Psi(t),
\end{equation}
where $\Psi(t)$ is the statevector of the system, and $H(t)$ is the Hamiltonian that incorporates the interaction with a radiation field. Typically, when a long laser pulse interacts with an atom or a simple molecule, e.g., nanosecond lasers, the interaction can be modeled by considering only the states coupled by the radiation. This approximation is justified because the bandwidth of the laser, that means, the uncertainty in energy of the laser photon, is inversely proportional to the pulse duration ($\Delta E \Delta \tau \geq 2\pi$). Thus, for long pulses, this bandwidth is significantly smaller compared to the energy difference between the system's states. 

\begin{figure}[ht!]
\begin{center}
\includegraphics[width=5cm]{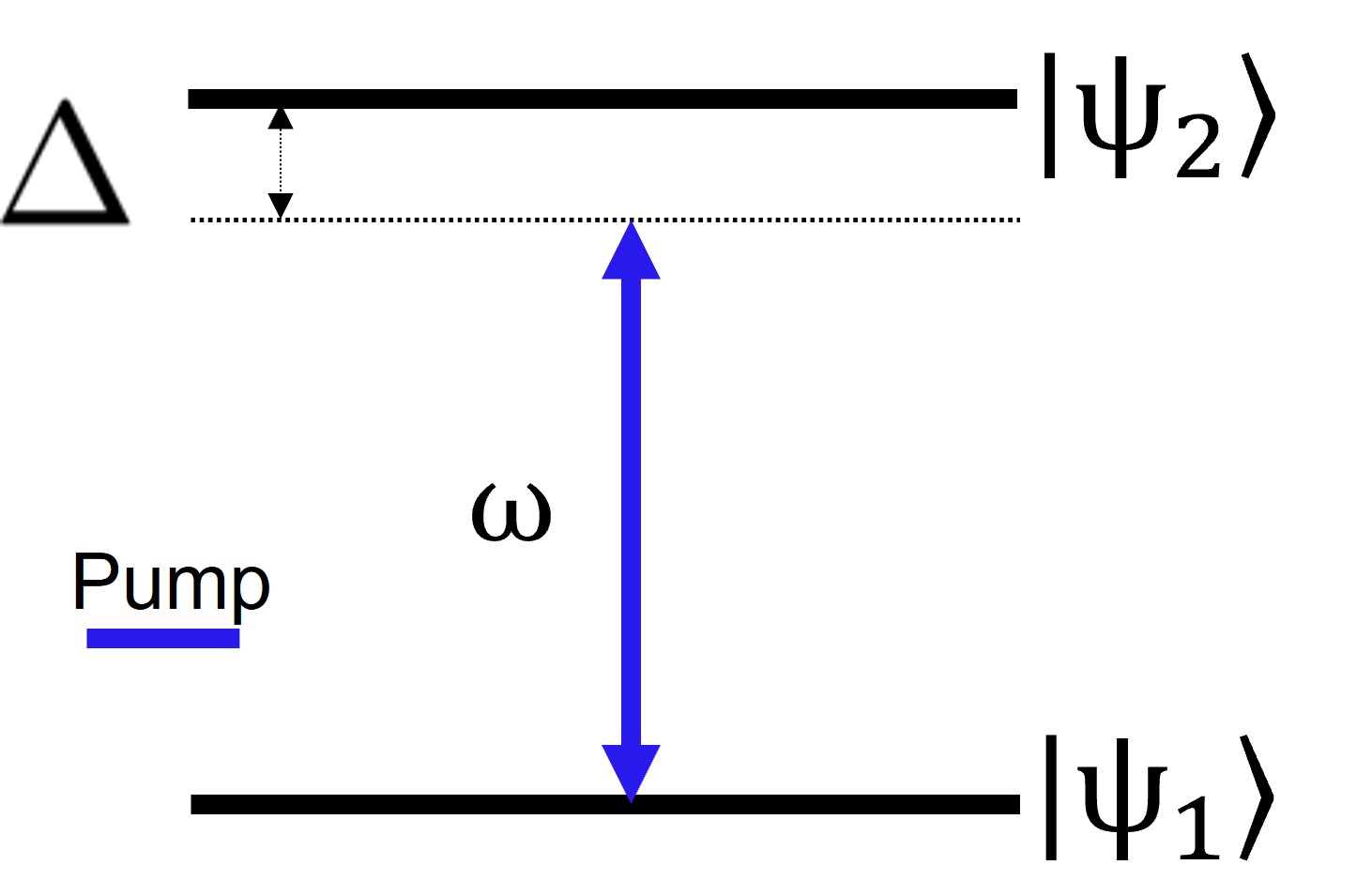}
\caption{\label{twolevelsytem} Two-state system interacting with a radiation field. The laser detuning from resonance is represented by $\Delta$. The energies of the excited and ground states are $E_2$ and $E_1$ respectively.}
\end{center}
\end{figure}

Let us consider a two-state system interacting with the electric field of a laser pulse with carrier frequency $\omega$, which is detuned from the transition frequency by $\Delta$. This situation is described by Fig\,\ref{twolevelsytem}.  The time dependent Hamiltonian in the electric dipole approximation can be written as
\begin{equation}
\label{hamidefi}
H(t)=H^0+H^I(t),
\end{equation}
where $H^0$ is the two-state field free Hamiltonian with energies $E_1$ and $E_2$
\begin{equation}
H^0= \left[
\begin{array}{cc}
 E_1 & 0 \\
 0 & E_2
\end{array}
\right],
\end{equation}
and $H^I$ is the interacting Hamiltonian
\begin{equation}
H^I(t)= \left[
\begin{array}{cc}
0  & \langle \psi_1
|V| \psi_2 \rangle   \\
 \langle \psi_2 |V| \psi_1 \rangle   & 0
\end{array}
\right],
\end{equation}
with
\begin{equation}
\label{laser_radiation}
V=-\textbf{d}\cdot[\textbf{e}\mathcal{E}(t)\cos \omega t],
\end{equation}
being $\textbf{d}$ the dipole moment, and $\textbf{e}$ and $\mathcal{E}(t)$ the polarization direction and the slow varying envelope of the electric field respectively. It is important to acknowledge that in the equations above we have neglected the contribution of the magnetic dipole to the interaction, as well as higher-order interacting terms. This approximation is justified by the significant difference in magnitude between electric and magnetic dipole moments, and it remains valid as long as laser radiation is not extremely intense  \cite{Sh90, Cohen, Sh08}. However, for ultraintense lasers, such as Petawatt lasers, it becomes necessary to consider higher-order interacting terms in order to accurately describe the interaction.

We can further simplify the Hamiltonian by applying the Rotating Wave Approximation (RWA) \cite{Sh90}. In Eq.\,\ref{laser_radiation}, we observe that the term $\mathcal{E}(t)$ represents the contribution of the slowly varying laser envelope, while the term $\omega$ corresponds to the fast oscillations of the laser frequency. Typically, our interest lies in studying the population dynamics over multiple optical cycles. Consequently, these fast oscillations can be effectively averaged out. The RWA provides an excellent approximation as long as the laser contains several optical cycles. However, for ultrashort laser pulses consisting of only one or two cycles, the RWA cannot be applied \cite{Sh90, Fleming10}. Thus, the RWA-Hamiltonian reads 
\begin{equation}
\label{ch2hamiltonian}
H^{RWA}=\frac{\hbar}{2}\left[
\begin{array}{cc}
0 & \Omega(t)\\
\Omega(t) & 2\Delta
\end{array}
\right],
\end{equation}
where $\Omega(t)$ is the Rabi frequency, i.e., the interaction energy divided by $\hbar$, 
\begin{equation}
\label{ch2definitonofrabi}
\Omega(t)=-(\textbf{d}\cdot\textbf{e})\mathcal{E}(t)/\hbar,
\end{equation}
and $\Delta$ the laser detuning
\begin{equation}
\Delta=\frac{(E_2-E_1)}{\hbar}-\omega.
\end{equation}

It is convenient to describe the population dynamics in the adiabatic basis \footnote{This basis is sometimes called dressed basis -dressed by the radiation- in contrast to the diabatic or bare basis.} $\left[\Phi_+, \Phi_-\right]$, which is the basis formed by the instantaneous eigenstates of the Hamiltonian described by Eq.\,\ref{ch2hamiltonian}. These adiabatic eigenstates are obtained by diagonalizing the RWA-Hamiltonian shown in Eq.\,\ref{ch2hamiltonian}. Thus, the eigenstates can be written as \cite{Vitanov01}:
\begin{equation}
\label{Phi+}
\Phi_+ (t)=\psi_1\sin \vartheta(t)+\psi_2 \cos\vartheta(t),
\end{equation}
\begin{equation}
\label{Phiminus} \Phi_- (t)=\psi_1\cos \vartheta(t)-\psi_2 \sin\vartheta(t),
\end{equation}
where $\vartheta(t)$ is the mixing angle 
\begin{equation}
\label{mixingangle}
\vartheta(t)=\frac{1}{2}\arctan\frac{\Omega(t)}{\Delta},
\end{equation}
and $\psi_j$ ($j=1,2$) are the bare states. The corresponding eigenenergies are
\begin{equation}
\lambda_\pm=\frac{1}{2}\left[\Delta\pm\sqrt{\Omega^2(t)+\Delta^2}\right].
\end{equation}
Thus, the statevector of the system in the basis formed by the adiabatic
states $\left[\Phi_+, \Phi_-\right]$, can be written as
\begin{equation}
\label{state_adiabatic}
\Psi(t)=B_+(t)\Phi_++B_-(t)\Phi_-.
\end{equation}
while in the diabatic basis  $\left[\psi_1,
\psi_2\right]$ it is written as
\begin{equation}
\label{state_diabatic}
\Psi(t)=C_1(t)\psi_1+C_2(t)\psi_2.
\end{equation}

The relation between the probability amplitude of the diabatic states, i.e., bare states, $\textbf{C}(t)=[C_1(t), C_2(t)]^T$, and the adiabatic states $\textbf{B}(t)=[B_+(t), B_-(t)]^T$, is
\begin{equation}
\label{relation_amplitude}
\textbf{C}(t)=R[\vartheta(t)]\textbf{B}(t),
\end{equation}
where
\begin{equation}
R[\vartheta(t)]=\left[
\begin{array}{cc}
\cos\vartheta & \sin\vartheta \\
-\sin\vartheta & \cos\vartheta
\end{array} \right].
\end{equation}

To express the time-dependent Schr\"{o}dinger equation (see Eq.\,\ref{timedepsch}) in the adiabatic basis, we substitute Eq.\,\ref{relation_amplitude} into Eq.\,\ref{timedepsch}. Taking into account that $R(\vartheta)R(-\vartheta)$ results in the identity matrix, we can write:
\begin{equation}
i\hbar\frac{d}{dt}\textbf{B}(t)=H_b(t)\textbf{B}(t),
\end{equation}
with
\begin{equation}
\label{adiabatichamiltonian}
 H_b=R(-\vartheta)HR(\vartheta)-i\hbar
R(-\vartheta)\frac{d}{dt}R(\vartheta)= \hbar\left[
\begin{array}{cc}
\lambda_- &
-i\dot{\vartheta}\\
i\dot{\vartheta}&
\lambda_+
\end{array}\right].
\end{equation}

Once the adiabatic Hamiltonian and the Schr\"{o}dinger equation in the adiabatic basis are defined, it is important to establish the conditions for adiabatic evolution. If the Hamiltonian (Eq.\ref{ch2hamiltonian}) varies sufficiently slowly in time, indicating adiabatic evolution, the statevector $\Psi(t)$ of the system will always remain aligned with a single adiabatic state or a coherent superposition of adiabatic states. In other words, if the initial statevector $\Psi(t)$ is aligned parallel to a specific adiabatic state, then during adiabatic evolution, $\Psi(t)$ will remain parallel to this particular state. This implies that the statevector $\Psi(t)$ will follow the temporal evolution of the adiabatic state, and there will be no mixing between different adiabatic states. Consequently, the non-diagonal terms in Eq.\,\ref{adiabatichamiltonian} can be neglected resulting in a diagonal Hamiltonian. This condition can be mathematically expressed as shown in Eq\,\ref{adiabatic_condition_eq} (see \ref{adiabatic_condition} for further details):
\begin{equation}
\label{adiabatic_condition_eq}
|\Delta|\gtrsim1/\tau
\end{equation}
being $\Delta$ the laser detuning and $\tau$ the laser pulse duration. It is important to notice that the adiabatic condition is independent of $\Omega(t)$, and therefore of the laser intensity.

\subsection{Coherent Population Return (CPR)}
Let us assume that a two state system interacting with a laser pulse with Rabi frequency  $\Omega(t)$ to be negligibly small outside a finite time interval $t_i < t < t_f$, i.e., outside the pulse duration $\tau = t_f-t_i$. Also, consider a situation where the laser frequency $\omega$ is detuned from exact resonance (see Fig.\,\ref{twolevelsytem}), being the adiabatic evolution condition defined previously fulfilled, i.e., $\Delta \gtrsim 1/\tau$. If at the beginning of the interaction ($t=-\infty$) all the population is in the ground state $\psi_1$, i.e., in Eq.\,\ref{state_diabatic} $C_2(-\infty)=0$, the statevector of the system $\Psi$ is aligned parallel to the adiabatic state $\Phi_-(t).$ 
\begin{equation}
\Psi(-\infty)=\psi_1=\Phi_-(-\infty)
\end{equation}
This is so because according to the definition of the adiabatic states (see Eq.\,\ref{Phi+} and Eq.\,\ref{Phiminus}) for $t=-\infty$ the mixing angle $\vartheta(-\infty)=0$ and, therefore, $\Phi_-(-\infty)=\psi_1$.

Since the adiabatic condition is fulfilled by hypothesis, the Hamiltonian varies sufficiently slow in time, and the state vector of the system $\Psi(t)$ remains always aligned with the adiabatic state $\Phi_-(t)$. Initially, at the beginning of the interaction ($t = -\infty$), when the Rabi frequency is zero ($\Omega(-\infty) = 0$) and the mixing angle is zero ($\vartheta(-\infty) = 0$), the population starts in the ground state. However, during the interaction ($t_i < t < t_f$), when the Rabi frequency is nonzero ($\Omega(t) \neq 0$) and the mixing angle is nonzero ($\vartheta \neq 0$), the state vector $\Psi(t)$ becomes a coherent superposition of the bare states $\psi_1$ and $\psi_2$, leading to transient excitation of the population to the upper state. Once the interaction has ceased ($t = +\infty$), when the Rabi frequency returns to zero ($\Omega(+\infty) = 0$) and the mixing angle goes back to zero ($\vartheta(+\infty) = 0$), the state vector $\Psi(t)$ becomes aligned with the ground state $\psi_1$ again. As result, any population that was transferred to the excited state during the interaction returns completely to the ground state.

\begin{figure}
\begin{center}
\includegraphics[width=\textwidth]{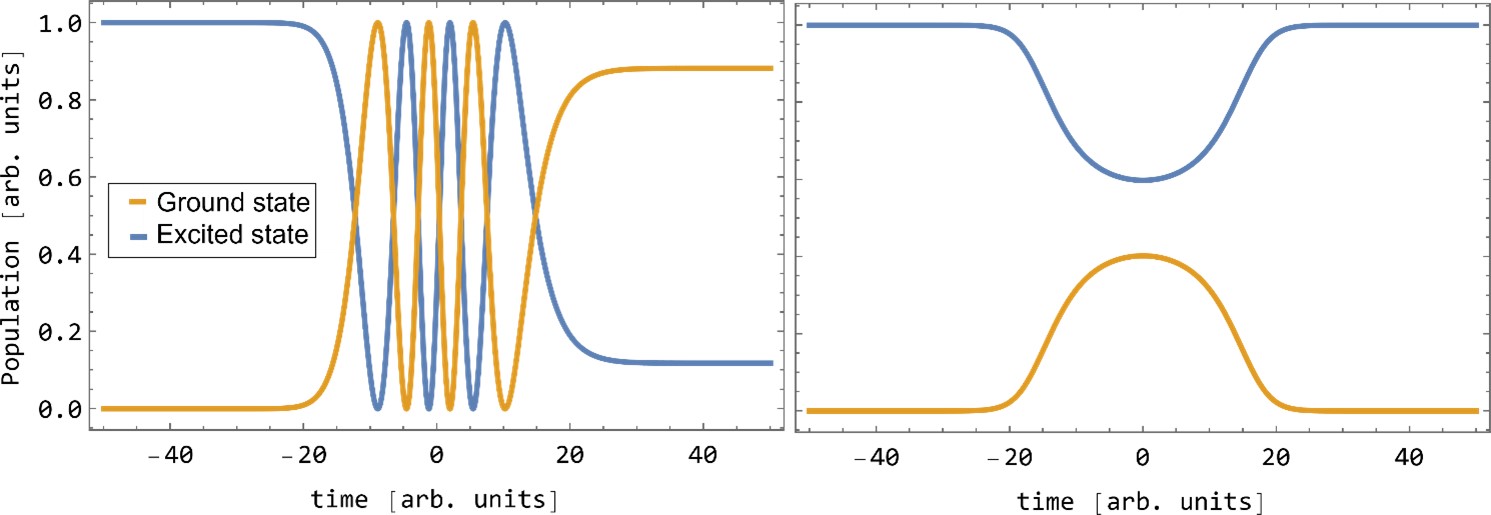}
\caption{\label{popubare} Population dynamics for a two level system on resonance (left frame) and when the adiabatic condition for CPR is satisfied ( $\Delta \gtrsim 1/\tau$) (right frame). In the first case, the population exhibits Rabi oscillations \cite{Sh90} while for CPR the population dynamics is smooth.}
\end{center}
\end{figure}

According to Eq.,\ref{Phiminus}, the population in the ground state ($P_1(t) = |C_1(t)|^2$) and the excited state ($P_2(t) = |C_2(t)|^2$) can be written as follows:
\begin{equation}
\label{Populations}
P_1(t)=\cos^2 \vartheta(t)\quad P_2(t)=\sin^2 \vartheta(t)
\end{equation}
It is important to emphasize that no population remains permanently in the excited state, regardless of the transient intensity of the laser pulse. This holds true even if the laser pulse intensity is very high (see Fig.\,\ref{popubare}). The adiabatic evolution ensures that the population transferred to the excited state during the interaction returns completely to the ground state once the interaction has ceased. 
This coherent phenomenon is known as Coherent Population Return (CPR) as has been subject of many theoretical and experimental studies \cite{Vitanov01, Peralta06, Peralta09, Peralta10, Chacon15, Boyero16, Vaquero17}.

\subsection{Introducing a probe laser}
Let us consider now a second laser (probe laser) to detect, e.g., by photoionization, the population transferred to the excited state $\psi_2$. We assume the probe laser intensity is sufficiently small so it can be treated like a small perturbation, and as a consequence there is no direct photoionization from the ground state, e.g., REMPI processes. According to our previous discussion, if the system evolves adiabatically ($|\Delta|\gtrsim1/\tau$) no population remains in the excited state once the pump pulse has ceased due to CPR, and the ionization signal is produced exclusively when pump and probe pulses overlap in time (see Fig.\,\ref{popubare}). On the other hand, if the evolution of the system is diabatic, meaning $|\Delta|\lesssim1/\tau$, there is a permanent population transferred to the excited state, and consequently, ionization signal will be produced even when pump and probe pulse do not temporally overlap. We should remember that we did not consider any relaxation channel from the excited state. 

\begin{figure}
\begin{center}
\includegraphics[width=5cm]{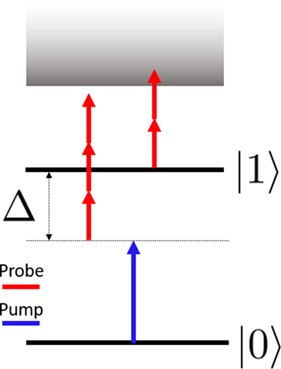}
\caption{\label{implementation} Possible experimental scheme.}
\end{center}
\end{figure}

The situation we have discussed is paradigmatic because according to CPR, it is possible the transfer of population from the ground state to an excited state, even when the energy of the pump laser photons is lower than the energy gap between the ground and excited states (as depicted in Fig-\,\ref{twolevelsytem}). Once the population is in the excited state, the energy of the probe laser can be adjusted to match the energy gap between the excited state and the continuum, and, therefore, the sum of the absorbed energies from the pump and probe lasers would be smaller than the energy required for ionizing the system.  Importantly, the detuning of the pump laser energy can be as large as desired, and the population will still be transferred to the excited state and subsequently ionized by the probe laser. This demonstrates the flexibility of the method, as the ionization process can be achieved even with significant detuning. Figure\,\ref{implementation} illustrates a possible experimental scheme, where nonresonant ionization requires the absorption of three photons, while CPR-mediated ionization only requires the absorption of two photons. 

The discussion so far has covered well-known aspects of CPR, but there has been one puzzling question that remained unanswered: how is energy conserved in this process? According to the results, the absorption of pump and probe photons provides enough energy to ionize the system, even though the sum of the absorbed photon energies is much smaller than the ionization potential (as shown in Fig.,\ref{implementation}). There seems to be a discrepancy in energy conservation, and this has been a source of confusion. However, CPR has been implemented in numerous experimental and theoretical studies, indicating that the concept is valid. Thus, there must be a misconception or missing piece in the theory presented so far. Our journey begins here, where we acknowledge that there have been mistakes and wrong ideas along the way because energy must be conserved in the process. In the following sections, we will present the different ideas we generated and discuss them in the order they were explored, leading us to the eventual solution.

\subsubsection{Nonresonant ionization} \label{nonresonant_Section}

The initial idea that came to mind was to consider the ionization signal as a result of the nonresonant absorption of pump and probe photons. In other words, instead of transiently promoting the population to an excited state, it was proposed that the population absorbs the necessary probe photons from a virtual state to reach the continuum (refer to Fig.,\ref{implementation}). However, three key arguments dismiss this possibility:

\begin{enumerate}

\item The current theoretical treatment of the time-dependent Schr\"{o}dinger equation does not account for the nonresonant ionization channel. To incorporate it, one would need to introduce it ad hoc, typically as a loss channel, in the time-dependent Hamiltonian outlined in Eq.\,\ref{ch2hamiltonian}.

\item As it is shown in Fig.\,\ref{implementation}, it is possible to tune the pump laser frequency to change the number of probe photons required to ionize the system from the resonant and the non-resonant situation. If the ionization order is different, and since the ionization signal is proportional to $I^n$ being $I$ the probe laser intensity and $n$ the number of photons absorbed, it would be possible to discriminate between both channels by recording the ionization signal versus the probe laser intensity. The display of these data in logarithmic-logarithmic plot must be a line with slope equal to the ionization order $n$. 

Furthermore, one could think that maybe two pump photons are absorbed in the process. This possibility does not hold because in  Eq.\,\ref{ch2hamiltonian} is not considered the multiphoton absorption. Also, according to Eq.\,\ref{Populations} the population transferred to the excited state by CPR can be written as
\begin{equation}
\label{popu_state_2}
P_2(t)=\sin\vartheta(t)=\frac{1}{2}-\frac{\Delta}{2\sqrt{\Omega^2(t)+\Delta^2}}.
\end{equation}
Under conditions of weak laser-matter interaction, specifically $\Omega(t)\tau\ll1$, it can be demonstrated \cite{Peralta09} that the population in the excited state can be approximated as:
\begin{equation}
\label{P2Aprox}
P_2(t)\simeq\frac{\Omega^2(t)}{\Delta^2}\propto I_{\textrm{Pump}(t)}.
\end{equation}
This equation indicates that even when the pump laser is significantly detuned from resonance, the transfer of population from the ground to the excited state by CPR is solely attributed to the absorption of a single pump laser photon.

\item If the ionization signal is indeed generated through the non-resonant absorption of pump and probe photons, where ionization from the excited state does not occur, it would require the pump and probe lasers to be synchronized in time. However, this contradicts the theoretical treatment of CPR outlined previously. Specifically, if the adiabatic condition (as shown in Eq.,\ref{adiabatic_condition_eq}) is not satisfied, meaning $|\Delta|\lesssim1/\tau$, there would be a permanent transfer of population to the excited state. This characteristic of CPR has been experimentally utilized to address various challenges, such as preventing power broadening \cite{Demtroder}—the broadening of homogeneous line profiles caused by intense laser fields—, in the detection of traces of isotopes by selective laser ionization \cite{Peralta06} or in the analysis of ultrafast decays mechanisms of complex molecules in time-resolved experiments \cite{Peralta10}.
\end{enumerate}

\subsubsection{Stark shift}
The raised doubts about the conservation of energy during the CPR process may appear to be explainable by Stark shifts, which have been extensively used for population manipulation of quantum systems through frequency-chirping techniques for many years (see, for example, \cite{Vitanov01_2} and references therein). In these schemes, the transition frequency is chirped or swept by the interaction with the electric field of the laser radiation through the laser frequency, instead of adjusting the laser frequency to match the transition frequency. However, it is important to consider that the magnitude of the Stark shift depends on the intensity of the radiation and the detuning from any off-resonant state. Thus, inducing significant shifts requires high laser intensities \cite{Sh90}. To be more specific, the dynamic Stark shift between an excited state $|\psi_2\rangle$ and a ground state $|\psi_1\rangle$, coupled to an off-resonant manifold of states $|\psi_i\rangle$, can be described by Eq.\,\ref{Stark_equ} \cite{Sh90}:
\begin{equation}
\label{Stark_equ}
S=-\frac{1}{4\hbar^2}\sum_i \left( \frac{|\textbf{d}_{2i} \cdot \textbf{e}|^2}{\Delta_{2i}}-\frac{|\textbf{d}_{1i} \cdot \textbf{e}|^2}{\Delta_{1i}} \right) E^2
\end{equation}
Thus, it was only with the development of nanosecond laser pulses that chirping techniques became a viable option for population manipulation (see, for example, \cite{Rickes_00, Rickes_03}). One of these techniques is SCRAP (Stark Shift Rapid Adiabatic Passage) \cite{Rickes_00}, where a pair of laser pulses, namely Stark shifting pulse and coupling pulse, interact with a system achieving population transfers of 100\%. In the case of CPR, in contrast to SCRAP, only one laser is involved. Nevertheless, one might speculate that the interaction of this laser with the system, even at low intensities, could modify the level structure through Stark shifts and sweep the transition through the CPR laser frequency. However, there are two key arguments that rule out this possibility:
\begin{itemize}
\item The theoretical formalism for CPR presented earlier does not account for Stark shifts. To consider this effect, the coupling to all off-resonance levels must be taken into account, requiring a term similar to Eq.\,\ref{Stark_equ} to be included in Eq.\,\ref{ch2hamiltonian}. However, this is not the case because we are restricting ourselves to a two-level system.

\item According to the theoretical description of CPR presented earlier, population is transferred from the ground state to the excited state for all pump laser intensities (see Eq.\,\ref{popu_state_2}). This contradicts the hypothesis of Stark shifts being responsible for the population transfer. If Stark shifts were the driving mechanism, one would expect that below a certain laser intensity threshold the electric field of the radiation would not be strong enough to sweep the resonance across the pump laser frequency. However, this behavior is not observed in the presented theoretical formalism.
\end{itemize}

\subsubsection{The concept of photon}

It is widely acknowledged that our current description of electromagnetic radiation, whether as a wave or as photons, is only applicable to non-interacting fields. Consequently, when the system's Hamiltonian depends on time, such as during interactions with laser pulses, the conservation of energy is not expected as explained, for example, in \cite{Cohen}. In Quantum Mechanics time and energy are conjugate variables related by an uncertainty principle meaning that is impossible to simultaneously determine both variables with arbitrary precision (see Eq.\,\ref{t_E}).
\begin{equation}
\label{t_E}
\Delta E\Delta t \geq \hbar/2
\end{equation}
The nature of time in quantum mechanics is indeed a fascinating and distinctive aspect that warrants careful analysis. Unlike other physical quantities such as position or momentum, time does not have an associated operator in the formalism of Quantum Mechanics. This fundamental difference raises intriguing questions and challenges our intuitive understanding of time in the quantum realm. While, in classical physics, time is commonly regarded as an absolute and universally flowing entity, in Quantum Mechanics the nature of time becomes more elusive and less straightforward due to the absence of an associated operator. This absence implies that time is treated differently, and it is not subject to the same direct measurements and manipulations as other observables like for example for momentum and position variables. As a consequence, the uncertainty relation presented earlier (see Eq. \ref{t_E}) does not have a direct derivation from the formalism of Quantum Mechanics, unlike the well-known position-momentum uncertainty relation. The position-momentum uncertainty relation arises naturally from the commutation relations between position and momentum operators. However, in the case of time, since it does not have an associated operator, the uncertainty relation involving time and energy (see Eq. \ref{t_E}) is not derived directly from the mathematical framework. The uncertainty relation involving time and energy is a result of theoretical considerations and physical arguments. 

The absence of a time operator raises intriguing questions and challenges our conventional understanding of time in the quantum domain. It means that time is not considered a fundamental observable that can be directly measured or used to extract information about a quantum system. Instead, time is often treated as an external parameter that governs the evolution of quantum states and processes. This topic warrants a more comprehensive analysis and, in line with the present manuscript, can be highly instructive for university-level students seeking to expand their understanding of Quantum Mechanics.

Based on the preceding discussion and in accordance with Eq. \ref{t_E}, it becomes apparent that during an interaction, the precise determination of the system's energy is not possible. This conclusion can also be reached by examining the equation that governs the time evolution of an observable represented with associated operator A, as presented in "Quantum Mechanics" by Cohen-Tannoudji (pg 241) \cite{Cohen}:
\begin{equation}
\label{evoper_prev}
 \frac{d\langle\Psi(t)|A|\Psi(t)\rangle}{dt}=\frac{1}{\imath\hbar}\langle\Psi(t)|[A,
H(t)]|\Psi(t)\rangle+\langle\Psi(t)|\frac{\partial A}{\partial
t}|\Psi(t)\rangle.
\end{equation}
This equation describes how the expectation value of the observable A changes with time. The right-hand side consists of two terms: the commutator of A with the Hamiltonian H(t) and the expectation value of the partial derivative of A with respect to time. Since the Hamiltonian is the associated operator to energy, the second term in the previous equation becomes zero
\begin{equation}
[H(t), H(t)]=0,
\end{equation}
obtaining finally
\begin{equation}
\label{Energy_operator}
 \frac{d\langle\Psi(t)|H|\Psi(t)\rangle}{dt}= \frac{dE}{dt}=\langle\Psi(t)|\frac{\partial H}{\partial
t}|\Psi(t)\rangle.
\end{equation}
 Thus, if the Hamiltonian explicitly depends on time, meaning that the expectation value of the time derivative of the Hamiltonian is non-zero, i.e., $\langle\Psi(t)|\frac{\partial H}{\partial
t}|\Psi(t)\rangle\neq 0$, then energy is not conserved in the system. This situation often occurs during an interaction process or when there are time-dependent external fields or potentials influencing the system. In such cases, the system's energy can change over time, and energy conservation does not hold. Hence, both from the analysis of Eq. \ref{t_E} and the equation governing the time evolution of observables, we arrive at the conclusion that the exact energy of the system cannot be precisely determined during interactions. According to this discussion, we can conclude that on one hand to achieve an accurate energy balance, it is necessary to consider an additional term, namely the interaction energy, and on the other hand, and as a consequence, during interactions involving a time-dependent Hamiltonian, the concept of a photon becomes ambiguous. In fact, it is worth noting that the concept of a photon originates from the quantization of a time-independent electromagnetic field.

However, and despite the previous discussion, at the very beginning of the interaction and once the pump pulse has ceased there is no longer a time-dependent interaction. Consequently, it becomes possible to reliably balance the energy of the system. Let's assume that at the end of the interaction, the system was ionized through the combined action of the pump and probe lasers. Since, as discussed earlier (see Section \ref{nonresonant_Section}), non-resonant ionization is not feasible, we must assume that ionization occurs via the excited state $|\psi_2\rangle$ being the energy of the system at the end of the interaction higher than at the beginning. This leads to another intriguing question: if the concept of a photon becomes blurred during the interaction, how many photons from the pump laser have been absorbed to promote the population from the ground state to the excited state during the CPR process? Is it possible to consider the absorption of a non-integer number of photons during the interaction?

At this point, we must admit that we were completely confused. It is possible that we had missed or misunderstood something, but coherent population redistribution (CPR) had been extensively studied and implemented theoretical and experimentally. Therefore, since our description had been so far semiclassical, with the electric field not being quantized, we decided to delve deeper into the analysis by employing second quantization of the electric field. This approach would enable us to effectively count the number of photons involved in the process.

\subsubsection[The Jaynes-Cummings model]{The Jaynes-Cummings model}
The theoretical description presented so far is based on a semiclassical approximation of the interaction between lasers and matter. In this approximation, matter is treated using the principles of Quantum Mechanics, while the laser radiation is described classically. Usually, this semiclassical approach provides an accurate description of the interaction between lasers and matter, and it is unnecessary to quantize the electric field of the radiation. However, we have enhanced the theoretical description presented above by incorporating the quantization of the electric field. This addition was motivated by our belief that effectively counting the number of photons involved in the process will provide a crucial element for understanding how energy conservation operates in CPR.

 In principle the quantization of the electric field is only necessary when working with close to single photon interaction processes \footnote{A typical nanosecond laser pulse of energy of 300\,$\mu$J and wavelength 400\,nm contains of the order of 10$^{14}$ photons.}, but such treatment allows us to count exactly the photon that are absorbed and emitted by the system, and to calculate exactly the energy of the system before, during and after the interaction.  For taking into account the quantization of the radiation, we have studied a slightly modified version of the Jaynes-Cummings model (JCM). This model was developed to study a two level atom interacting with a quantized mode of an optical cavity, with or without the presence of light (see for example \cite{Knight93} or the issue commemorating the 50 years of the Jaynes-Cumming physics and references therein  \cite{Greentree13}).

\begin{figure}[ht!]
\begin{center}
\includegraphics[width=9cm]{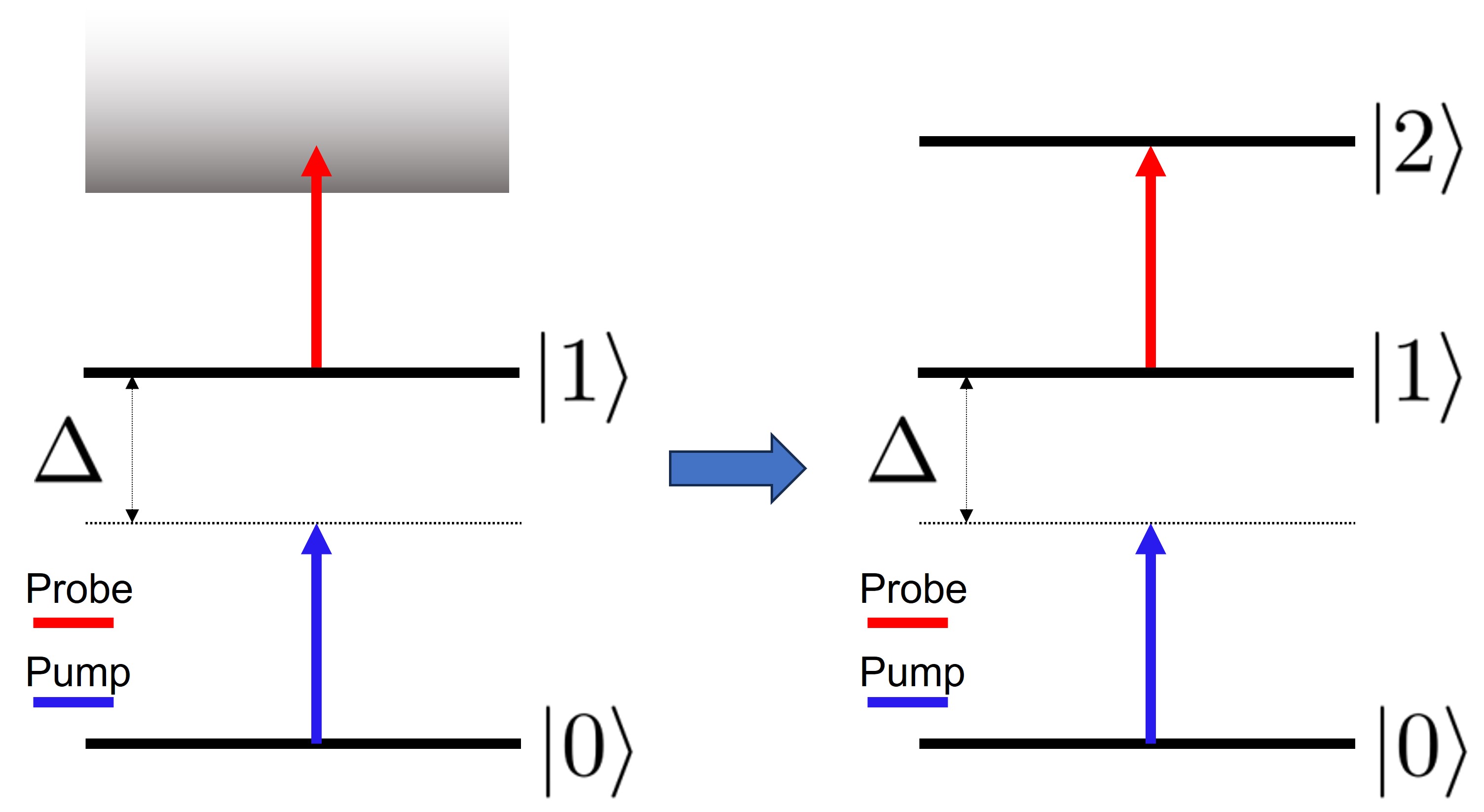}
\caption{\label{JCM-model} The continuum is
modeled as another bound state.}
\end{center}
\end{figure}

The right part of Fig.\,\ref{JCM-model} illustrates the system we modeled using the Jaynes-Cummings formalism. Initially, we considered a pump laser that is detuned from resonance, resulting in the transient population transfer via CPR from the ground state to the excited state. The population transferred to the latter state is probed by a second laser, i.e., by a probe laser. Since we treated the continuum as a loss channel, without incorporating a bound state within the unstructured continuum through a dressing interaction (as described in, for example, \cite{Dalton90} and references therein), we could replace the interaction with the continuum by considering the interaction with a second dressed state, denoted as $|2\rangle$. This substitution is valid as long as the intensity of the probe laser remains low. The latter condition justifies treating the interaction between states $|1\rangle$ and $|2\rangle$ as a perturbation. 

There were two main reasons for adopting this approach. Firstly, incorporating the continuum into the Schr\"{o}dinger equation is a challenging task that requires further approximations. Secondly, the three-level system depicted in Fig.\,\ref{JCM-model} is a closed system, enabling a straightforward calculation of the system's energy and the number of photons.
The Hamiltonian in the JCM formalism of the system showed in the right part of  Fig.\,\ref{JCM-model}, it is composed of three distinct parts:
\begin{equation}
H^{JCM}=H_{atom}+H_{light}+H_{interaction}
\end{equation}

\begin{itemize}
\item The Hamiltonian of the atom $H_{atom}$ can be written as
\begin{equation}
H_{atom}=0\sigma_0+\omega_{01}\sigma_1+\omega_{02}\sigma_2
\end{equation}
being
\begin{equation}
\sigma_0=\begin{pmatrix}
1 & 0 & 0  \\
0 & 0 & 0   \\
0 & 0 & 0
\end{pmatrix}
\hspace{1cm} \sigma_1=\begin{pmatrix}
0 & 0 & 0  \\
0 & 1 & 0   \\
0 & 0 & 0
\end{pmatrix}
\hspace{1cm} \sigma_2=\begin{pmatrix}
0 & 0 & 0  \\
0 & 0 & 0   \\
0 & 0 & 1
\end{pmatrix}
\end{equation}
being 0, $\omega_{01}$, and $\omega_{02}$ the energy of the state $|0\rangle$, $|1\rangle$, and $|2\rangle$ respectively. For the sake of simplicity, we have omitted the use of $\hbar$ in this section.

\item The Hamiltonian of the light $H_{light}$  can be written as
\begin{equation}
H_{light}=\omega_{L1} (aa^+)+\omega_{L2}(bb^+)
\end{equation}
where $\omega_{L1}$ and $\omega_{L2}$ correspond to the frequency of the pump and probe laser respectively, and a, a$^+$, b, and b$^+$ are the creation and annihilation operators of pump and probe laser photons respectively \cite{Cohen}. For simplicity, we have omitted the zero energy of the electric fields. 

\item The Hamiltonian of the interaction $H_{interaction}$
\begin{equation}
H_{interaction}=\omega_1(a^+\sigma_{1-} + a\sigma_{1+}
+a^+\sigma_{1+} +a\sigma_{1-})+ \omega_2(b^+\sigma_{2-} +
b\sigma_{2+} +b^+\sigma_{2+} +b\sigma_{2-})
\end{equation}
where $\omega_1$ and $\omega_2$ are the Rabi frequency per photon corresponding to pump and probe laser respectively, and 
\begin{equation}
\sigma_{1-}=\begin{pmatrix}
0 & 1 & 0  \\
0 & 0 & 0   \\
0 & 0 & 0
\end{pmatrix}
\hspace{1cm} \sigma_{1+}=\begin{pmatrix}
0 & 0 & 0  \\
1 & 0 & 0   \\
0 & 0 & 0
\end{pmatrix}
\hspace{1cm} \sigma_{2-}=\begin{pmatrix}
0 & 0 & 0  \\
0 & 0 & 1   \\
0 & 0 & 0
\end{pmatrix}
\hspace{1cm} \sigma_{2+}=\begin{pmatrix}
0 & 0 & 0  \\
0& 0 & 0   \\
0 & 1 & 0
\end{pmatrix}
\end{equation}
It is important to remark that no additional approximation, beyond the dipole approximation and the Rotating Wave Approximation (RWA) is considered. 
\end{itemize}

According to the previous discussion, the Hamiltonian without interaction H$_{0}$ is defined as $H_{0}=H_{atom}+H_{light}$, and the total Hamiltonian $H=H_{0}+H_{Interaction}$. Furthermore, since we are now considering the photon numbers of pump and probe lasers, the description basis cannot remain as simple as $\{|0\rangle, |1\rangle, |2\rangle\}$, becoming necessary to work in the following basis $|atom\,state, n, m\rangle$ where $n$ represents the number of pump photons and $m$ represents the number of probe photons. This extended basis allows us to account for the varying photon numbers in the pump and probe lasers.

The chosen basis is indeed infinite, so it is necessary to truncate it. It is important to note that due to the structure of the interaction Hamiltonian, which includes terms that involve the excitation of an atom by emitting a photon and the relaxation by absorbing a photon, the working basis must be sufficiently large. For this purpose, the following basis has been selected:
\begin{equation}
\begin{aligned}
B=\{|0, n, m\rangle, |1, n-1, m\rangle, |2, n-1, m-1\rangle, |2, n-1, m+1\rangle, |1, n-1, m+2\rangle, |1, n+1, m\rangle, \\ |2, n+1, m-1\rangle, |2, n+1,  m+1\rangle, |1, n+1, m+2\rangle, |0, n+2,
m\rangle \}
\end{aligned}
\end{equation}
This selection ensures that the basis is sufficiently large to account for the relevant interactions involving the emission and absorption of photons by the atoms in the system. After a tedious work, H in the basis B reads:
\begin{equation}
\label{matrixH}
\tiny
\rotatebox{-90}{$\textrm{H}= \left(\begin{matrix}
n\omega_{L_1}+m \omega_{L_2} & \omega_1 \sqrt{n} & 0 & 0 & 0 \\
\omega_1 \sqrt{n} & \Delta_1+n \omega_{L_1}+m \omega_{L_2} & \omega_2 \sqrt{m} & \omega_2 \sqrt{m+1} & 0\\
0 &\omega_2 \sqrt{m}  & \Delta_1+\Delta_2+n \omega_{L_1}+m \omega_{L_2} &0 & 0  \\
0 &\omega_2 \sqrt{m+1}  & 0 &\Delta_1+\Delta_2+n \omega_{L_1}+(m+2) \omega_{L_2} & \omega_2 \sqrt{m+2}\\
0 & 0  & 0 &\omega_2 \sqrt{m+2}& \Delta_1+n\omega_{L_1}+(m+2)\omega_{L_2}\\
\omega_1\sqrt{n+1} & 0  & 0 &0 &0 \\
0 & 0  & 0 &0 &0 \\
0 & 0  & 0 &0 &0 \\
0 & 0  & 0 &0 &0 \\
0 & 0  & 0 &0 &0\\
.... & .... &.... &.... & ....\\
\omega_1 \sqrt{n+1} & 0 & 0 & 0 & 0 \\
0 & 0 & 0 & 0 & 0\\
0 & 0 & 0 & 0 & 0\\
0 & 0 & 0 & 0 & 0\\
0 & 0 & 0 & 0 & 0\\
\Delta_1+(n+2)\omega_{L_1}+m\omega_{L_2}& \omega_2\sqrt{m}& \omega_2\sqrt{m+1}& 0 & \omega_1\sqrt{n+2}\\
\omega_2\sqrt{m} &\Delta_1+\Delta_2+(n+2)\omega_{L_1}+m\omega_{L_2}& 0&0&0\\
\omega_2\sqrt{m+1} & 0 & \Delta_1+\Delta_2+(n+2)\omega_{L_1}+(m+2)\omega_{L_2}& \omega_2\sqrt{m+2} &0\\
0 & 0 & \omega_2\sqrt{m+2} & \Delta_1+(n+2)\omega_{L1}+(m+2)\omega_{L2}& 0 \\
\omega_1 \sqrt{n+2}& 0 & 0 &0 &(n+2)\omega_1+m\omega_2
\end{matrix}\right)$}
\end{equation}
where
\begin{equation}
\Delta_1=\omega_{01}-\omega_{L_1}\hspace{2cm}\Delta_2=(\omega_{02}-\omega_{01})-\omega_{L_2}=\omega_{12}-\omega_{L_2}.
\end{equation}

Once we have the Hamiltonian H expressed in the basis B, we can proceed to solve the time-dependent Schr\"{o}dinger equation, being $|\Psi(t)\rangle$ the instantaneous eigenvector of the system, i.e.,
 \begin{equation}
 \begin{aligned}
 |\Psi(t)\rangle=C_1(t)|0, n, m\rangle+C_2(t)|1, n-1, m\rangle+C_3(t)|2, n-1, m-1\rangle+C_4(t)|2, n-1,
m+1\rangle\\+C_5(t)|1, n-1, m+2\rangle+C_6(t)|1, n+1,
m\rangle+C_7(t)|2, n+1, m-1\rangle+\\C_8(t)|2, n+1,
m+1\rangle+C_9(t)|1, n+1, m+2\rangle+C_{10}(t)|0, n+2, m\rangle
 \end{aligned}
\end{equation}
with initial conditions
\begin{equation}
\begin{aligned}
\{C_1(t),C_2(t),C_3(t),C_4(t),C_5(t),C_6(t),C_7(t),C_8(t),C_9(t),C_{10}(t)\}_{t=0}=\{1,0,0,0,0,0,0,0,0,0\},
\end{aligned}
\end{equation}
and compute the total population in states $\{|0\rangle, |1\rangle,
|2\rangle\}$, i.e.,
\begin{equation}
\begin{aligned}
P_0(t)=C_1(t)C_1^*(t)+C_{10}(t)C_{10}^*(t)\\
P_1(t)=C_2(t)C_2^*(t)+C_5(t)C_5^*(t)+C_6(t)C_6^*(t)+C_9(t)C_9^*(t)\\
P_2(t)=C_3(t)C_3^*(t)+C_4(t)C_4^*(t)+C_7(t)C_7^*(t)+C_8(t)C_8^*(t),
\end{aligned}
\end{equation}
the instantaneous mean energy of the system, 
\begin{equation}
E(t)=\langle\Psi(t)|H|\Psi(t)\rangle,
\end{equation}
the number of photons of the pump laser
\begin{equation}
\begin{aligned}
N=nP_1(t) + (n - 1)P_2(t) + (n - 1)P_3(t) + (n - 1)P_4(t)+ (n -
1)P_5(t)\\ +  (n + 1)P_6(t) + (n + 1)P_7(t) + (n + 1)P_8(t) + (n +
1)P_9(t) + (n + 2)P_{10}(t),
\end{aligned}
\end{equation}
the number of photons of the probe laser
\begin{equation}
\begin{aligned}
M=mP_1(t) + mP_2(t) + (m - 1)P_3(t) + (m +1)P_4(t)+ (m +2)P_5(t)\\
+  mP_6(t) + (m -1)P_7(t) + (m + 1)P_8(t) + (m + 2)P_9(t) +
mP_{10}(t),
\end{aligned}
\end{equation}
and the excitation number
\begin{equation}
\begin{aligned}
ExcNumber(t)=ExcNumber_{01}(t)+ExcNumber_{12}(t)=\\((1/2)+(1/2)(P_2(t)+P_1(t)-P_0(t))+N)
+((1/2)+(1/2)(P_2(t)-P_1(t)-P_0(t))+M).
\end{aligned}
\end{equation}

For simplicity, we can define the quantities
\begin{equation}
Inver_{01}=(1/2)+(1/2)(P_2(t)+P_1(t)-P_0(t))\hspace{1cm}Inver_{12}=(1/2)+(1/2)(P_2(t)-P_1(t)-P_0(t)),
\end{equation}
resulting
\begin{equation}
ExcNumber(t)=ExcNumber_{01}(t)+ExcNumber_{12}(t)=Inver_{01}+N+Inver_{12}+M.
\end{equation}
 The excitation number takes into the number of photons into the system and the number of atomic excitations. The sum of both numbers must remain constant, i.e., for each absorbed photon an excitation must be produced and viceversa (see for example \cite{Sh90} pg 630). 

\begin{figure}[ht!]
\begin{center}
\includegraphics[width=\textwidth]{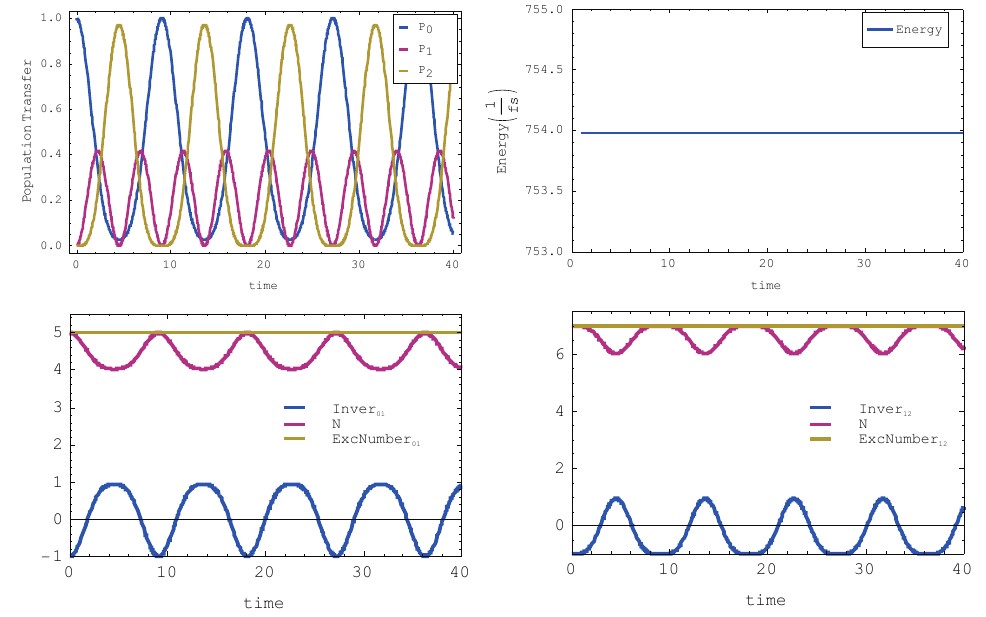}
\caption{\label{simunotime} Populations, energy, ExcNumber$_{01}$, and ExcNumber$_{12}$. The parameters of 
the simulation are: $\omega_{L_1}=20\pi\,\textrm{fs}^{-1}$, $\omega_{L_2}=20\pi\,\textrm{fs}^{-1}$, $\Delta_1=0$, $\Delta_2=0$,
$\omega_1=0.2\,\textrm{fs}^{-1}$, $\omega_2=0.2\,\textrm{fs}^{-1}$,
$N_{t=0}=5$, $M_{t=0}=7$}
\end{center}
\end{figure}

\begin{figure}[ht!]
\begin{center}
\includegraphics[width=\textwidth]{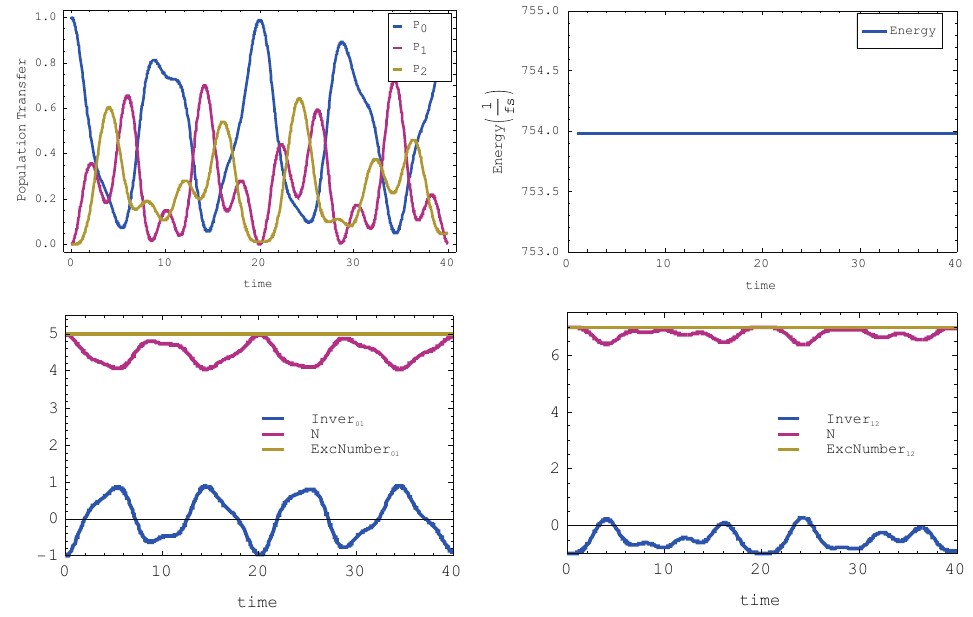}
\caption{\label{simunotimedetu}
Populations, energy, ExcNumber$_{01}$, and ExcNumber$_{12}$. The parameters of the simulation are: $\omega_{L_1}=20\pi\,\textrm{fs}^{-1}$, $\omega_{L_2}=20\pi\,\textrm{fs}^{-1}$, $\Delta_1=0.2\pi\,\textrm{fs}^{-1}$, $\Delta_2=0$, $\omega_1=0.2\,\textrm{fs}^{-1}$, $\omega_2=0.2\,\textrm{fs}^{-1}$,
$N_{t=0}=5$, $M_{t=0}=7$}
\end{center}
\end{figure}

Let's consider the simplest case where the atom is irradiated by a continuous wave (CW) laser, meaning that the Hamiltonian of the system does not depend on time. In this situation, as shown in Eq.\,\ref{Energy_operator} the energy remains constant. Figure\,\ref{simunotime} shows the solution for the time-dependent Schr\"{o}dinger equation with the Hamiltonian described by Eq.\,\ref{matrixH} for an interaction that does not depend on time.
As expected, while the populations change over time, the energy and the excitation number (ExcNumber) remain constant. The constant excitation number throughout the time evolution of the system dynamics indicates that photons are absorbed as whole units, maintaining their integer nature. In simpler terms, this addresses the speculation raised in the previous section and confirms that the absorption of non-integer photons is not possible. This finding underscores the discrete and quantized nature of photon absorption within the system. This observation holds true even if $\Delta_1\neq0$ (see Fig. \ref{simunotimedetu}).

Let us now consider pulsed excitation. We introduce time dependence in the Rabi frequencies per photon, denoted as $\omega_1$ and $\omega_2$, while keeping the number of photons (N and M) constant. It is important to note, that in experiments with pulsed lasers, the number of photons changes with time, resulting in a varying Rabi frequency. This scenario can be understood as a cavity that allows only one electromagnetic mode inside it, filled with atoms and photons, where the length of the cavity changes with time. The latter situation must be analogous to a situation where the excitation is pulsed. 

\begin{figure}[ht!]
\begin{center}
\includegraphics[width=\textwidth]{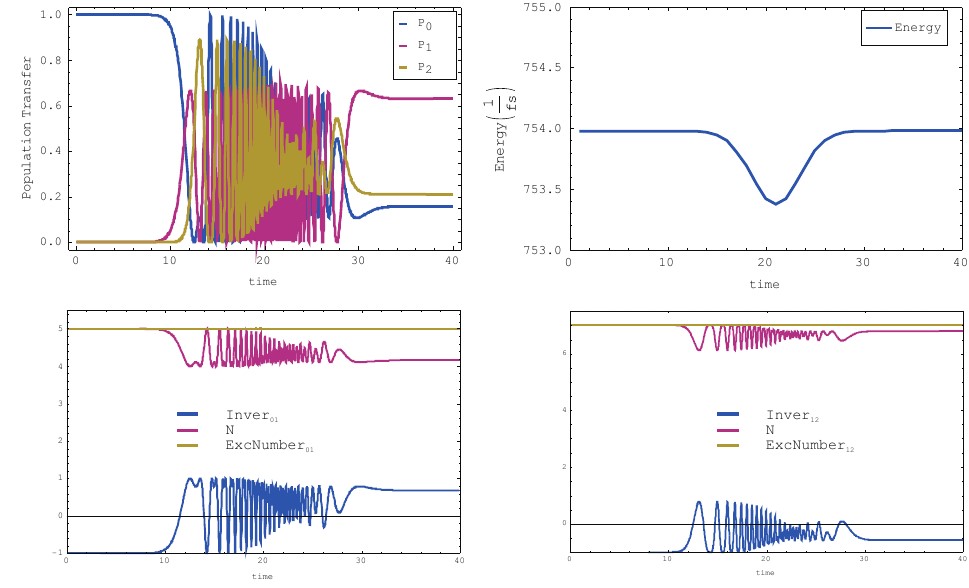}
\caption{\label{simutime} Populations, energy, ExcNumber$_{01}$, and ExcNumber$_{12}$. The parameters of
the simulation are: $\omega_{L_1}=20\pi \textrm{fs}^{-1}$, $\omega_{L_2}=20\pi\,\textrm{fs}^{-1}$, $\Delta_1=0$, $\Delta_2=0$, $\omega_1=5\exp[-((t - 20)^2/25)]\,\textrm{fs}^{-1}$, $\omega_2=5\exp[-((t - 20)^2/25)]\,\textrm{fs}^{-1}$, $N_{t=0}=5$,
$M_{t=0}=7$}
\end{center}
\end{figure}

Let us study first the resonant case, i.e., the frequency of the pump laser matches exactly the Bohr frequency between states $|0\rangle$ and $|1\rangle$ the energy (see Fig.\,\ref{simutime}). 
As in the previous examples, the excitation number (ExcNumber) remains constant throughout the process. However, since the interaction depends on time, the energy of the system, as given by Eq.\,\ref{evoper}, is not constant during the interaction. Nonetheless, we can compare the energy of the system before the interaction (at $t=-\infty$) with the energy after the interaction (at $t=+\infty$) because $H_{interaction}=0$ in both cases. Therefore, we have:
\begin{equation}
H=H_0+H_{Interaction}(t)|_{t=\pm\infty}=\textrm{H}_0
\end{equation}
being the Hamiltonian time independent. As shown in Fig. \ref{simutime}, the energy at $t=-\infty$ is equal to the energy at $t=+\infty$.

\begin{figure}[ht!]
\begin{center}
\includegraphics[width=\textwidth]{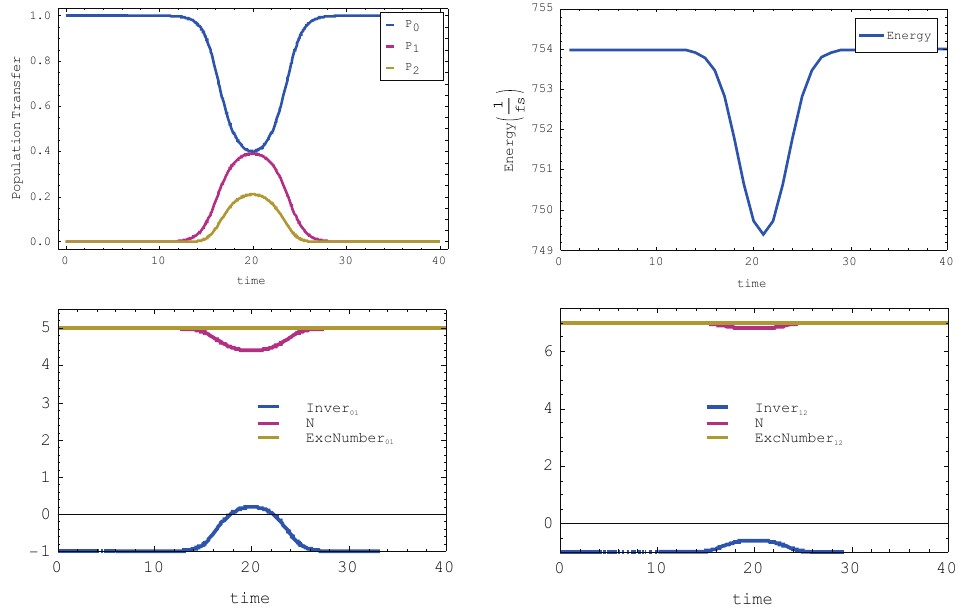}
\caption{\label{simuCPR1} Populations, energy, ExcNumber$_{01}$, and ExcNumber$_{12}$. The parameters of
the simulation are: $\omega_{L_1}=20\pi \textrm{fs}^{-1}$, $\omega_{L_2}=20\pi\,\textrm{fs}^{-1}$, $\Delta_1=2\pi\,\textrm{fs}^{-1}$, $\Delta_2=0$,
$\omega_1=2\exp[-((t - 20)^2/25)]\,\textrm{fs}^{-1}$, $\omega_2=3\exp[-((t - 20)^2/25)]\,\textrm{fs}^{-1}$, $N_{t=0}=5$, $M_{t=0}=7$}
\end{center}
\end{figure}

After obtaining the previous results, which served as a simulation test, the next step was to consider a time dependent excitation but with $\Delta_1\neq0$, i.e., the CPR case. Running the simulation for different parameters, we observed two distinct situations. If the parameters were chosen in a way that no population remained in states $|2\rangle$ and/or $|3\rangle$ after the excitation, the energy before and after the excitation remains equal (see Fig. \ref{simuCPR1}). It must be noticed that during the dynamics it appears that a dual CPR process is occurring: one between $|0\rangle$ and $|1\rangle$ induced by the pump laser and another between $|1\rangle$ and $|2\rangle$ induced by the probe laser.

\begin{figure}[ht!]
\begin{center}
\includegraphics[width=\textwidth]{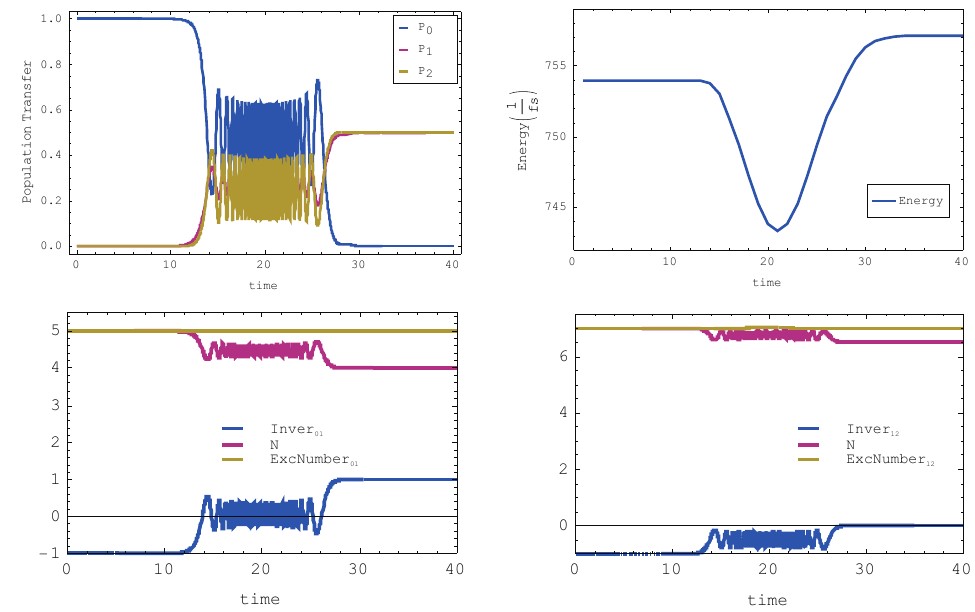}
\caption{\label{simuCPR2} Populations, energy, ExcNumber$_{01}$, and ExcNumber$_{12}$. The parameters of the simulation are: $\omega_{L_1}=20\pi \textrm{fs}^{-1}$, $\omega_{L_2}=20\pi\,\textrm{fs}^{-1}$, $\Delta_1=\pi\,\textrm{fs}^{-1}$, $\Delta_2=0$, $\omega_1=2\exp[-((t- 20)^2/25)]\,\textrm{fs}^{-1}$, $\omega_2=8\exp[-((t -20)^2/25)]\,\textrm{fs}^{-1}$, $N_{t=0}=5$, $M_{t=0}=7$}
\end{center}
\end{figure}

On the other hand, if the parameters were chosen is such a way that some population remained in the states $|1\rangle$ and/or $|2\rangle$ after the excitation, the energy before and after the excitation did not not remain constant (see Fig.\,\ref{simuCPR2}).
The simulations shown in Fig.\,\ref{simuCPR2} might suggest that the system, comprising atoms and light in the Jaynes-Cummings model, is capable of gaining energy "seemingly out of nothing." However, in reality, the energy gained per excited atom, once the excitation has ceased exactly corresponds to the detuning of the pump laser from resonance. This mathematical relationship can be expressed as follows: 
\begin{equation}
\Delta_1=\frac{E(+\infty)-E(-\infty)}{P_1(+\infty)+P_2(+\infty)}
\end{equation}
This equation shows that the gained energy is calculated by considering the difference in energy between the final state at $t=+\infty$ and the initial state at $t=-\infty$, divided by the sum of the populations of the relevant states at $t=+\infty$. Based on the results and observations, it appears that the Jaynes-Cummings formalism developed in this section may not provide a complete explanation regarding how energy is conserved in a coherent population return (CPR) process. 

At this moment, we found ourselves perplexed, and realized that it might be necessary to expand our theoretical framework and incorporate additional factors and interactions that influence the energy dynamics of the system. In other words, maybe we needed to move beyond the limitations of the presented model. This expansion could encompass exploring the intricacies of atom-photon coupling dynamics, understanding the impact of external fields, and delving into the overall energy balance within the system. However, by a stroke of luck or serendipity-, we stumbled upon an idea that provided the key element to solve the paradox. As is often the case, the solution turned out to be much simpler than the extensive efforts we had undertaken to explain the apparent paradox.

\section[Solution]{Solution}

\begin{figure}
\includegraphics[width=4cm]{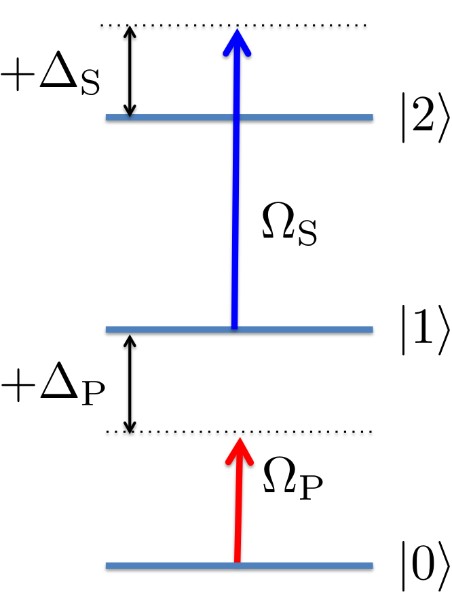}
\centering
\caption{\label{fig10} (Color online) Level scheme. It is important to note the sign convention for the detunings.}
\end{figure}

The key insight that solved the paradox involved replacing the probing photoionization process with a bound-bound transition. By doing so, the continuum was eliminated, allowing for easy calculation of the system's energy. Consider a three-level system as depicted in Fig.\,\ref{fig10}, where a pump laser, detuned by $\Delta_P$ from resonance, couples the states $|0\rangle$ and $|1\rangle$. Additionally, a second laser, known as the Stokes laser, couples the states $|1\rangle$ and $|2\rangle$. In our description, we make the following assumptions:
\begin{itemize}
\item The detuning $\Delta_P$ is greater than the inverse of the pulse duration  of the pump laser $\tau_P$, i.e., the laser bandwidth. This ensures that the adiabatic condition for CPR between states $|0\rangle$ and $|1\rangle$ is satisfied.

\item The perturbation induced by the Stokes laser is very small compared to that of the pump laser ($\Omega_P \gg \Omega_S$), placing us in the context of of "weak measurement” or “non-perturbative measurement”.

\item The temporal duration of the pump pulse, $\tau_P$, is significantly longer than that of the probe pulse, $\tau_S$ ($\rm \tau_P>\tau_S$).
\end{itemize}

The Hamiltonian that describe the situation depicted in Fig.\,\ref{fig10} in the Rotating Wave Approximation (RWA) reads (see for example \cite{Sh90, Sh11} for the mathematical treatment):
\begin{equation}
\label{Hamilt_STIRAP}
 H(t)=\frac{\hbar}{2}\left(
\begin{array}{ccc}
0 &  \Omega_P(t) & 0\\
\Omega_P(t) & 2\Delta_P &  \Omega_S(t) \\
 0 &  \Omega_S(t) & 2(\Delta_P-\Delta_S)
\end{array}
\right),
\end{equation}
Since we assume that the effect of the Stokes laser is negligible compared to the dynamics induced by the pump laser, it becomes possible to separate the Hamiltonian defined in Eq.\,\ref{Hamilt_STIRAP} into blocks. Specifically, we can isolate the block that accounts for the interaction between states $|0\rangle$ and $|1\rangle$. This block can be represented as:
\begin{equation}
\label{Hamilt_STIRAP_2}
\hat{H}_{\{0,1\}}(t)=\frac{\hbar}{2}\left(
\begin{array}{cc}
0 & \Omega_P(t) \\
\Omega_P(t) & 2\Delta_P
\end{array}
\right).
\end{equation}

By diagonalizing this block independently from state $|2\rangle$, we can analyze it separately. This Hamiltonian, denoted as $\hat{H}_{\{0,1\}}$, corresponds to the typical coherent population return (CPR) scheme in a two-level system, as discussed in the previous sections. The adiabatic (dressed) eigenstates of $\hat{H}_{\{0,1\}}$ are given by:
\begin{equation}
\label{phiminus_ap}
\begin{aligned}
 |\Phi_- (t)\rangle=\cos \vartheta(t)|0\rangle- \sin \vartheta(t)|1\rangle\\
 |\Phi_+ (t)\rangle=\sin \vartheta(t)|0\rangle+\cos\vartheta(t)|1\rangle
\end{aligned}
\end{equation}
with associated eigenvalues $\left(\lambda_-(t), \lambda_+(t)\right)$
\begin{equation}
\label{lambdas}
\lambda_\mp(t)=\frac{1}{2}\left[\Delta_P\mp\sqrt{\Omega_P^2(t)+\Delta_P^2}\right]
\end{equation}
where the mixing angle is defined by
\begin{equation}
\vartheta(t)=(1/2)\arctan[\Omega_P(t)/\Delta_P].
\end{equation}

In the adiabatic basis formed by the states $B_{Adiab.}=\{|\Phi_-(t)\rangle, |\Phi_+(t)\rangle, |\psi_2\rangle\}$ the hamiltonian of Eq.\,\ref{Hamilt_STIRAP} -Hamiltonian written in the diabatic basis $B_{Diab.}=\{|\psi_0\rangle, |\psi_1\rangle,|\psi_2\rangle\}$- reads (see \ref{Hamilt_adiab} for the mathematical details):
\begin{equation}
\label{Hamilt_STIRAP_adiabatic}
H_{Adib.}(t)=\hbar\left(
\begin{array}{ccc}
\lambda_- & 0 & -\frac{1}{2}\Omega_S(t)\sin\vartheta(t) \\
 0 &  \lambda_+ &  \frac{1}{2}\Omega_S(t)\cos\vartheta(t) \\
 -\frac{1}{2}\Omega_S(t)\sin\vartheta(t) &   \frac{1}{2}\Omega_S(t)\cos\vartheta(t) & (\Delta_P-\Delta_S)
\end{array}
\right).
\end{equation}

\begin{figure}
\includegraphics[width=8cm]{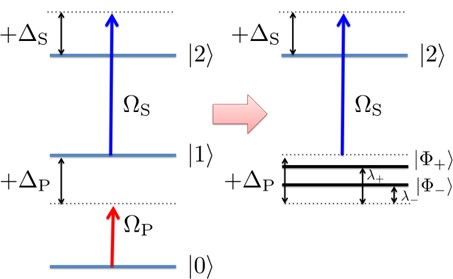}
\centering
\caption{\label{fig2_a} (Color online) Level scheme in the diabatic $B_{diab.}=\{|\psi_0\rangle, |\psi_1\rangle,|\psi_2\rangle\}$ and adiabatic $B_{Adiab.}=\{|\Phi_-(t)\rangle, |\Phi_+(t)\rangle, |\psi_2\rangle\}$ basis.}
\end{figure}

\begin{figure}
\center
\includegraphics[width=4cm]{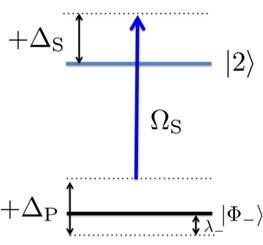}
\caption{\label{fig12} (Color online) Simplified level scheme in the adiabatic basis.}
\end{figure}

The level scheme in this new basis is shown in Fig.\,\ref{fig2_a}.
Also, since we assume that $\Delta_P > 1/\tau_P$, implying adiabatic evolution of the system when interacting with the pump laser $\Omega_P$, and initially all the population resides in the ground state, it follows from our previous discussion on CPR that the adiabatic state $|\Phi_+\rangle$ remains unpopulated. Consequently, we can simplify the Hamiltonian of Eq.\,\ref{Hamilt_STIRAP_adiabatic} to
\begin{equation}
\label{Hamilt_STIRAP_adiabatic_2}
H_{Adib.}'(t)=\hbar\left(
\begin{array}{cc}
\lambda_- &  -\frac{1}{2}\Omega_S(t)\sin\vartheta(t) \\
  -\frac{1}{2}\Omega_S(t)\sin\vartheta(t) & (\Delta_P-\Delta_S)
\end{array}
\right),
\end{equation}
and the level scheme depicted in Fig.\,\ref{fig2_a} to the one shown in Fig.\,\ref{fig12} 

By examining Eq.\,\ref{Hamilt_STIRAP_adiabatic_2}, it becomes evident that in order to populate state  $|2\rangle$, the detuning of the Stokes laser must be equal to (resonance condition): 
\begin{equation}
\begin{aligned}
\lambda_-=\Delta_P-\Delta_S\\
\Delta_S=\Delta_P-\lambda_-\\
\Delta_S=\frac{1}{2}\left[\Delta_P+\sqrt{\Omega_P^2(t)+\Delta_P^2}\right]
\end{aligned}
\end{equation}

\begin{figure}
\centering
\includegraphics[width=\textwidth]{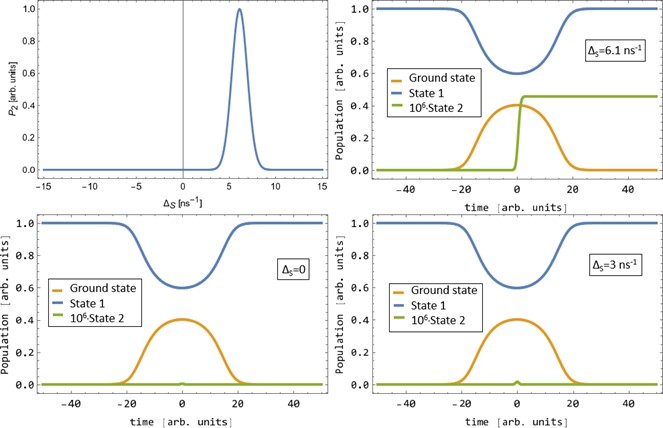}
\caption{\label{fig13} (Color online) 
Numerical results are obtained based on the level scheme illustrated in Fig.\,\ref{fig10}. The upper left frame displays the population of state $|2\rangle$ as a function of the detuning of the probe laser. The remaining frames depict the population dynamics for various detunings of the probe laser. It is noteworthy that for $\Delta_S=0$ and $\Delta_S=3\,\text{ns}^{-1}$, the population of state $|2\rangle$ exhibits transient dynamics characteristic of CPR induced by the interaction of two lasers, indicating that the probe laser is clearly off resonance. However, when $\Delta_S=6.1\,\text{ns}^{-1}$, the probe laser satisfies the resonance condition between states $|\Phi_-\rangle$ and $|2\rangle$, leading to a permanent transfer of population.}
\end{figure}

The obtained condition reveals an interesting insight: although one might assume that during the CPR process the Stokes laser couples the states  $|1\rangle$ and $|2\rangle$, it, in fact, couples states $|\Phi_-\rangle$ and $|2\rangle$. Given that the energy of state $|\Phi_-\rangle$ is lower than the energy of $|2\rangle$ (see Eq.\,\ref{lambdas}), this energy difference is compensated by the absorption of a more energetic Stokes laser photon, thereby preserving the conservation of energy principle. This theoretical result is confirmed by numerical simulations (see Fig.\,\ref{fig13}). The simulation was based on the scheme shown in Fig.\,\ref{fig10} and the following data:
\begin{itemize}
\item $\rm \Omega_P =\Omega_{0P}Exp[-(4 Ln[2])\frac{t^2}{\tau_P^2}];$ $\rm \Omega_S=\Omega_{0S}Exp[-(4 Ln[2])\frac{t^2}{\tau_S^2}]$
\item $\rm \Omega_{0P}=10\,ns^{-1}$; $\rm \Omega_{0S}=0.001\,ns^{-1}$
\item $\rm \tau_P=20\,ns$; $\rm \tau_S=2\,ns^{-1}$
\item $\rm \Delta_P=2\,ns^{-1}.$
\end{itemize}
According to our previous analysis the Stokes laser will be on resonance for a detuning $\rm \Delta_S$ equals to:
\begin{equation}
\rm \Delta_S=\frac{1}{2}\left[\Delta_P+\sqrt{\rm\Omega_P^2(t)+\Delta_P^2}\right]=6.1\,ns^{-1}.
\end{equation}
This agrees perfectly with the numerical simulation see Fig.\,\ref{fig13}

In conclusion, the mistake we made in interpreting the CPR dynamics was assuming that we could reason in terms of bare states during the process. In reality, this is not the case. As we discussed earlier, during CPR, the state that becomes populated is $|\Phi_-\rangle$, which has a lower energy than state $|1\rangle$. Consequently, any probing mechanism of the population transfer, such as photoionization with a Stokes laser, must take this energy difference into account to preserve the principle of energy conservation. In other words, the system's interaction with the lasers modifies the energy levels, underscoring the importance of considering the dynamics and energies of the adiabatic states rather than the bare states when describing laser-matter interaction processes. While it is common to discuss systems in terms of bare states, paying attention to finer details of the system dynamics can reveal the necessity for a different perspective.

It is worth noting that in the theoretical treatments conducted so far, both in the semiclassical approach and the Jaynes- Cummings formalism, this interaction energy has not been taken into account. In fact, in the latter formalism, a closed system is assumed, where the absorption and emission of photons can be precisely counted. However, as we have demonstrated, there are situations where energy is not conserved between the initial and final moments once the interaction has ceased. Furthermore, it is important to emphasize that this energy transfer, although reminiscent of Stark shifts, involves a distinct mechanism. On one hand, the magnitude of the shift in CPR is proportional to the Rabi frequency $\Omega(t)$, whereas for Stark shifts, it scales with the intensity of the laser. On the other hand, Stark shifts result from an off-resonance interaction between the laser and the other states of the system, requiring, therefor, the consideration of all those states. However, in the case of CPR, as we have previously shown, the involvement of any off-resonance state is not required. This energy shift can occur in a two-level system, highlighting the unique nature of the CPR process.

\section{Conclusion}
In conclusion, this work has presented how a seemingly simple question arising from a well-established theory, such as the semiclassical description of laser-matter interaction, can challenge our understanding of a fundamental principle of Physics: energy conservation. Along our journey to resolve this apparent paradox, we have revisited fundamental concepts, gaining a more detailed understanding of the intriguing hidden aspects encompassed by the theory of Quantum Mechanics. In fact, we believe that the value of reaching a definitive answer lies not only in the answer itself but also in the path taken to obtain it. This undoubtedly holds educational significance, aligning with our intention to stimulate comprehension of the phenomenon. By describing the process followed, we aspire to facilitate the understanding of this subject for students, while simultaneously deepening our own knowledge of laser-matter interaction and the contra-intuitive facts that Quantum Mechanics still comprises.  

\section*{Acknowledgments}
The author thank A. Longarte, R. Montero and G. Muga for the most valuable and stimulating discussions. 

\begin{appendices}
\section{The adiabatic condition}\label{adiabatic_condition}
For an adiabatic evolution, it is necessary that the energy splitting between the adiabatic eigenenergies $\lambda_+$ and $\lambda_-$ is significantly larger than the magnitudes of the non-diagonal elements in the Hamiltonian (Eq.\,\ref{adiabatichamiltonian}). This condition ensures that the Hamiltonian can be approximated as a diagonal matrix. The specific criterion for adiabaticity can be expressed as follows \cite{Vitanov01}:
\begin{equation}
\label{adiabaticevolution}
|\dot{\vartheta}(t)|\ll\lambda_+(t)-\lambda_-(t),
\end{equation}
or explicitly
\begin{equation}
|\Delta\dot{\Omega}(t)|\ll2[\Omega^2(t)+\Delta^2]^{3/2}.
\end{equation}
Defining the function
\begin{equation}
\label{ffunction}
f(t)=\frac{\Delta\dot{\Omega}(t)}{2[\Omega^2(t)+\Delta^2]^{3/2}}.
\end{equation}
the previous condition can be written as $f(t)\ll1$.

Since $f(t)$ is inversely proportional to $\Omega(t)$ and directly proportional to $\dot{\Omega}(t)$, the maximum of the function $f(t)$ should occur at early or late times during the laser pulse. This is because when the pulse reaches its maximum, i.e., when $\Omega(t)$ is maximum, $\dot{\Omega}(t)$ is equal to zero, resulting in $f(t)=0$. 

In experimental settings, laser pulses can often be reasonably modeled by slowly varying functions such as Gaussian or hyperbolic secant functions. Considering the hyperbolic secant function as an example, the temporal dependence of the Rabi frequency can be written as $\Omega(t) = \Omega_0 \textrm{sech}(t/\tau)$, where $\tau$ represents the pulse duration. In the limit of $t\rightarrow\infty$, we can approximate $\dot{\Omega}(t)$ as follows:
\begin{equation}
|\dot{\Omega}(t)|\approx\frac{\Omega(t)}{\tau}. 
\end{equation}
This approximation can be substituted in Eq.\,\ref{ffunction} obtaining
\begin{equation}
\label{ffunction_2}
f(t)=\frac{\Delta\Omega(t)}{2\tau[\Omega^2(t)+\Delta^2]^{3/2}}.
\end{equation}

Once obtained Eq.\,\ref{ffunction_2}, we can study where it presents a maximum as function of $\Delta$. Thus, 
\begin{equation}
\frac{\partial f(t)}{\partial \Delta}=\frac{2\Omega(t)\tau[\Omega^2(t)+\Delta^2]^{3/2}-6\Omega(t)\Delta^2\tau[\Omega^2(t)+\Delta^2]^{1/2}}{4\tau^2[\Omega^2(t)+\Delta^2]^{3}}=0,
\end{equation}
from where
\begin{equation}
[\Omega^2(t)+\Delta^2]^{2}=3\Delta^2,
\end{equation}
obtaining
\begin{equation}
\Omega(t)=\pm \sqrt{2}\Delta.
\end{equation}
Substituting the obtained value for the maximum in Eq.\,\ref{ffunction_2}, we finally obtain
\begin{equation}
\label{newadiab}
|f(t)|_{max}=\frac{1}{3\sqrt{6}|\Delta|\tau}.
\end{equation} 
According to our previous discussion for adiabatic evolution is required that $f(t)\ll1$ meaning that 
\begin{equation}
\label{newadiab}
|\Delta \tau|\gtrsim1,~\textrm{i.e.},~|\Delta|\gtrsim\frac{1}{\tau}.
\end{equation}
If this condition $|\Delta|\gtrsim\frac{1}{\tau}$ is not fulfilled the evolution will be diabatic. 

It is important to note that the adiabatic region is solely determined by the laser detuning parameter $\Delta$ and is independent of the peak Rabi frequency $\Omega_0$ (as shown in Eq.\,\ref{newadiab}). This detuning parameter can typically be easily controlled and adjusted in experiments. Furthermore, for achieving adiabatic evolution, a specific detuning value is not required; it is sufficient to satisfy Equation \ref{newadiab}. This simplifies the experimental implementation significantly. Additionally, it should be mentioned that for non-exponential pulse shapes, such as Gaussian pulses, there can be a slight dependence of the Rabi frequency on Eq.\,\ref{newadiab}. However, for smooth laser pulses, this dependence is typically negligible and does not significantly affect the adiabaticity condition.

\section{Hamiltonian expression in the adiabatic basis}\label{Hamilt_adiab}
The Hamitonian given by Eq.\,\ref{Hamilt_STIRAP_appendix} is expressed in the diabatic -bare- basis, denoted as $B_{diab.}=\{|\psi_0\rangle, |\psi_1\rangle, |\psi_2\rangle\}$
\begin{equation}
\label{Hamilt_STIRAP_appendix}
H_{B_{diab.}}(t)=\frac{\hbar}{2}\left(
\begin{array}{ccc}
0 & \Omega_P(t) & 0\\
\Omega_P(t) &  2\Delta_P & \Omega_S(t) \\
 0 &  \Omega_S(t) & 2(\Delta_P-\Delta_S)
\end{array}
\right).
\end{equation}
To express this Hamiltonian in the adibatic basis formed by $B_{Adiab.}=\{|\Phi_-(t)\rangle, |\Phi_+(t)\rangle, |\psi_2\rangle\}$, the following calculation is necessary:

\begin{equation}
\label{Hamilt_STIRAP_adiabatic_calculation}
H_{B_{Adib.}}(t)=\left(
\begin{array}{ccc}
 \langle \Phi_-^*|H_{B_{diab.}} |\Phi_-\rangle  & \langle \Phi_-^*|H_{B_{diab.}} |\Phi_+\rangle  & \langle \Phi_-^*|H_{B_{diab.}} |\psi_2\rangle \\
 \langle \Phi_+^*|H_{B_{diab.}} |\Phi_-\rangle & \langle \Phi_+^*|H_{B_{diab.}} |\Phi_+\rangle &  \langle \Phi_+^*|H_{B_{diab.}} |\psi_2\rangle \\
  \langle \psi_2^*|H_{B_{diab.}} |\Phi_-\rangle &  \langle \psi_2^*|H_{B_{diab.}} |\Phi_+\rangle & \langle \psi_2^*|H_{B_{diab.}} |\psi_2\rangle
\end{array}
\right).
\end{equation}
The first to terms in the main diagonal are easily calculated since $|\Phi_-\rangle$ and $|\Phi_+\rangle$ are by construction normalized eigenvectors of $H_{B_{diab.}}$, thus:
\begin{equation}
\langle \Phi_\mp^*|H_{B_{diab.}} |\Phi_\mp\rangle=\lambda_\mp \langle \Phi_\mp^*|\Phi_\mp\rangle=\lambda_\mp.
\end{equation}
The third term in the main diagonal has the same expression in both basis since the third vector, i.e., $|\psi_2\rangle$ is the same for both basis. The term $\langle \Phi_-^*|H_{B_{diab.}} |\Phi_+\rangle$ is directly zero since  $|\Phi_-\rangle$ and $|\Phi_+\rangle$ are by construction orthogonal vectors
\begin{equation}
\langle \Phi_-^*|H_{B_{diab.}} |\Phi_+\rangle=\lambda_+\langle \Phi_-^*|\Phi_+\rangle.
\end{equation}
The same reasoning works for  $\langle \Phi_+^*|H_{B_{diab.}} |\Phi_-\rangle$.

As an example, we demonstrate the calculation for one of the remaining terms:
\begin{equation}
\langle \Phi_-^*|H_{B_{diab.}} |\psi_2\rangle= |\Phi_- (t)\rangle=(\cos \vartheta(t)\langle0|- \sin \vartheta(t)\langle1|)H_{B_{diab.}}|\psi_2\rangle=- \sin \vartheta(t)\Omega_S(t)
\end{equation}
because states $\psi_0$ and $\psi_2$ are not connected but $\psi_1$ and $\psi_2$ are connected by the Stokes laser. Therefore, the Hamiltonian in the adiabatic basis is given by Eq.\,\ref{Hamilt_STIRAP_adiabatic_appendix}:
\begin{equation}
\label{Hamilt_STIRAP_adiabatic_appendix}
H_{Adib.}(t)=\hbar\left(
\begin{array}{ccc}
\lambda_- & 0 & -\frac{1}{2}\Omega_S(t)\sin\vartheta(t) \\
 0 & \lambda_+ &  \frac{1}{2}\Omega_S(t)\cos\vartheta(t) \\
 -\frac{1}{2}\Omega_S(t)\sin\vartheta(t) &  \frac{1}{2}\Omega_S(t)\cos\vartheta(t) & (\Delta_P-\Delta_S)
\end{array}
\right).
\end{equation}
\end{appendices}

\section*{Data availability statement}
No data associated in the manuscript.

\end{document}